\documentstyle[preprint,aps]{revtex}
\begin{document}
\draft
\title{Wigner's $D$-matrix elements for $SU(3)$ - A Generating Function 
Approach}
\author{J. S. Prakash{\thanks {jsp@iopb.ernet.in}}} 
\address{Institute Of Physics\\ Sachivalaya Marg, Bhubaneswar 751 005, India}
\date{January 1996}
\maketitle

\begin{abstract} 
A generating function for the Wigner's $D$-matrix elements of
$SU(3)$ is derived.  From this an explicit expression for the
individual matrix elements is obtained in a closed form.
\end{abstract}

PACS number(s): 

\section{Introduction}
The Wigner's $D$-matrix elements of $SU(3)$ have very important
applications in nuclear physics, particle physics, $SU(3)$
lattice gauge theories, matrix models, finite temparature field
theory calculations involving $SU(3)$ and other areas of
physics.  Starting with Murnaghan {\cite{mfd}}, who parametrized
the defining matrices of $U(n)$ and $O(n)$, many authors
{\cite{cemm,ntj,hdf}} have obtained expressions
for the Wigner's $D$-matrix elements of $SU(3)$ using various
methods.  It is the purpose of this paper to evaluate these
matrix elements for $SU(3)$ using the calculus we {\cite{jsphss}}
have set up to deal with computations involving the group
$SU(3)$.  The distinct advantage of this calculus and the
novelty of our present method is that it allows one to write a
generating function for these matrix elements from which one can
extract the individual matrix elements by using the auxiliary
inner product of the calculus.  

The plan of the paper is as follows. We begin, in section 2, by
reviewing the main ingredients of our calculus for $SU(3)$ which
are relevant to our present discussion and then, in section 3,
give a derivation of the generating function for the matrix
elements.  In section 4 we show how to extract the individual
matrix elements and obtain a polynomial expression for the
matrix elements in any irreducible representation in terms of
the matrix elements of the defining representation of $SU(3)$ in
any parametrization.  Section 5 is devoted to a discussion of
our results.  A few examples are included in the appendix for
illustrating the method.

\section{Overview of our previous results}

In this section we briefly review the results that we need on
the group $SU(3)$.  Some of these results were obtained by us in
a previous paper {\cite{jsphss}}.

$SU(3)$ is the group of $3\times 3$ unitary unimodular matrices
$A$ with complex coefficients. It is a group of $8$ real
parameters.  The matrix elements satisfy the following
conditions

\begin{eqnarray}
A&=&(a_{ij}), \nonumber \\ A^\dagger A &=& I, \qquad AA^\dagger
= I\, ,\,\,\, \mbox{where}\, I\, \mbox{is the identity matrix and\, ,}\nonumber \\
\mbox{det}(A)&=&1\, .
\label{A}
\end{eqnarray}

\subsection{Parametrization}

One well known parametrization of $SU(3)$ is due to
Murnaghan {\cite{mfd}}, see also {\cite{bmarh,cemm,ntj,bakd}}.
In this we write a typical element of $SU(3)$ as :

\begin{equation}
D(\delta_1,\delta_2,\phi_3) U_{23}(\phi_2,\sigma_3) U_{12}(\theta_1,\sigma_2) 
U_{13}(\phi_1,\sigma_1)\, ,
\end{equation}
with the condition $\phi_3 = -(\delta_1 + \delta_2)$.  Here $D$
is a diagonal matrix whose elements are ${\exp}(i\delta_1)$,
${\exp}(i\delta_2)$ ,${\exp}(i\phi_3)$ and $U_{pq}(\phi,\sigma)$ is a $3
\times 3$ unitary unimodular matrix which for instance in the
case $p=1$, $q=2$ has the form
\begin{equation}
\pmatrix{{\cos}\phi               & -{\sin}\phi {\exp}(-i\sigma) & 0 \cr
         {\sin} \phi {\exp}(i\sigma) & {\cos} \phi               & 0 \cr
         0                     & 0                      & 1}\, .
\end{equation}
The $3$ parameters $\phi_1$, $\phi_2$, $\phi_3$ are longitudinal
angles whose range is $-\pi \leq \phi_i \leq \pi$, and the
remaining $6$ parametrs are latitude angles whose range is
$\-\frac{1}{2} \pi \leq \sigma_i \leq \frac{1}{2}\pi$.

Now the trnasformations $U_{23}$ and $U_{13}$ can be changed into
transformations of the type $U_{12}$ whose matrix elements are
known, by the following device
\begin{eqnarray}
U_{13}(\phi_1, \sigma_1) &=& (2,3) U_{12}(\phi_1,\sigma_1) (2,3)\, , \nonumber \\
U_{23}(\phi_2, \sigma_3) &=& (1,2) (2,3) U_{12}(\phi_2,\sigma_3) (2,3) (1,2)\, ,
\end{eqnarray}
where $(1,2)$ and $(2,3)$ are the transposition matrices
\begin{equation}
(1,2) = \pmatrix{0 & 1 & 0 \cr
                 1 & 0 & 0 \cr
                 0 & 0 & 1 }, \hspace{.1in} (2,3) = \pmatrix{0 & 1 & 0 \cr
                                                      1 & 0 & 0 \cr
                                                      0 & 0 & 1 }\, .
\end{equation}

In this way the expression for an element of the $SU(3)$ group becomes
\begin{equation}
D(\delta_1,\delta_2,\phi_3) (1,2) (2,3) U_{12}(\phi_2,\sigma_3)
(2,3) (1,2) U_{12}(\theta_1,\sigma_2) (2,3)
U_{12}(\phi_1,\sigma_1) (2,3)\, .
\end{equation}

\subsection{Irreducible Representations.}

The above parametrization provides us with a defining
irreducible representation $\underline{3}$ of $SU(3)$ acting on
a $3$ dimensional complex vector space spanned by the triplet
$z_1, z_2, z_3$ of complex variables.  The hermitian adjoint of
the above matrix gives us another defining but inequivalent
irreducible representaion $\underline{3^*}$ of $SU(3)$ acting on
the triplet ${w_1,w_2,w_3}$ of complex variables spanning
another $3$ dimensional complex vector space.  Tensors
constructed out of these two $3$ dimensional representations
span an infinite dimensional complex vector space.  

\subsection{{{The Constraint}}}

If we impose the constraint

\begin{eqnarray}
 z_1w_1+z_2w_2+z_3w_3=0\, ,
\label{z.w}
\end{eqnarray}
on this space we obtain an infinite dimensional complex vector
space in which each irreducible representation of $SU(3)$ occurs
once and only once.  Such a space is called a model space for
$SU(3)$.  Further if we solve the constraint
$z_1w_1+z_2w_2+z_3w_3=0$ and eliminate one of the variables, say
$w_3$, in terms of the other five variables $z_1, z_2, z_3, w_1,
w_2$ we can write a genarating function to generate all the
basis states of all the IRs of $SU(3)$.  This generating
function is computationally a very convenient realization of the
basis of the model space of $SU(3)$.  Moreover we can define a
scalar product on this space by choosing one of the variables,
say $z_3$, to be a planar rotor ${\exp}(i\theta)$.  Thus the
model space for $SU(3)$ is now a Hilbert space with
this('auxiliary') scalar product between the basis states.  The
above construction was carried out in detail in a previous paper
by us {\cite{jsphss}}.  For easy accessability we give a
self-contained summary of those results here.

\subsection{Labels for the basis states.}

\noindent {{\bf{(i). Gelfand-Zetlein labels}}}\\

Normalized basis vectors are denoted by,
$\vert{M,N;P,Q,R,S,U,V}>$.  All labels are non-negative
integers.  All Irreducible Represenatations(IRs) are uniquely
labeled by $(M, N)$.  For a given IR $(M, N)$, labels
$(P,Q,R,S,U,V)$ take all non-negative integral values subject to
the constraints:

\begin{equation}
R+U=M\hspace{.1in},\hspace{.1in} S+V=N\hspace{.1in},\hspace{.1in} P+Q=R+S.
\end{equation}

The allowed values can be presribed easily: $R$ takes all values
from $0$ to $M$, and $S$ from $0$ to $N$.  For a given $R$ and
$S$, $Q$ takes all values from $0$ to $R+S$.\\

\noindent {{\bf{(ii). Quark model labels}}}.\\

The relation between the above Gelfand-Zetlein labels and the
Quark Model labels is as given below.

\begin{eqnarray}
2I=P+Q=R+S, \,\, 2I_3=P-Q, \,\, Y = \frac{1}{3} (M-N) + V-U 
= \frac{2}{3} (N-M)-(S-R)\, .\nonumber\\
\end{eqnarray}
where $R$ takes all values from $0$ to $M$.  $S$ takes all
values from $0$ to $N$.  For a given $R$ and $S$, $Q$ takes all
values from $0$ to $R+S$.
\subsection{Explicit realization of the basis states}
 
\noindent {\bf{(i). {Generating function for the basis states of $SU(3)$}}}

The generating function for the basis states of the IR's of
$SU(3)$ can be written as
 
\begin{equation}
g(p,q,r,s,u,v)={\exp}(r(pz_1+qz_2)+s(pw_2-qw_1)+uz_3+vw_3)\, .
\label{g()}
\end{equation}

The coefficient of the monomial $p^Pq^Qr^Rs^Su^Uv^V$ in the
Taylor expansion of Eq.(\ref{g()}), after eliminating $w_3$
using Eq.(\ref{z.w}), in terms of these monomials gives the
basis state of $SU(3)$ labelled by the quantum numbers $P, Q, R,
S, U, V$.

\noindent {\bf{{ (ii). Formal generating function for the basis
states of $SU(3)$ }}}

The generating function Eq.(\ref{g()}) can be written formally
as 

\begin{equation}
g=\sum_{P,Q,R,S,U,V} p^Pq^Qr^Rs^Su^Uv^V \vert PQRSUV)\, .
\label{FGF}
\end{equation}
where $\vert PQRSTUV)$ is an unnormalized basis state of $SU(3)$
labelled by the quantum numbers $P,Q,R,S,U,V$.

Note that the constraint $P+Q=R+S$ is automatically satisfied in
the formal as well as explicit Taylor expansion of the generating
function.

\noindent {\bf{{ (iii). Generalized generating function for the basis
states of $SU(3)$}}}

It is useful, while computing the normalizations(see below) of the basis
states, to write the above generating function in the following
form

\begin{equation}
{\cal G}(p,q,r,s,u,v)={\exp}(r_pz_1+r_qz_2+s_pw_2+s_qw_1+uz_3+vw_3)\, .
\label{GGF}
\end{equation}

In the above generalized generating function (\ref{GGF}) the
following notation holds.

\begin{equation}
r_p=rp, \qquad r_q=rq, \qquad s_p=sp, \qquad s_q=-sq\, .
\label{rprqspsq}
\end{equation}

\subsection{Notation}

Hereafter, for simplicity in notation we assume all variables
other than the $z^i_j$ and $w^i_j$ where $i, j=1,,2,3$ are real
eventhough we have treated them as comlex variables at some
places.  Our results are valid even without this restriction as
we are interested only in the coefficients of the monomials in
these real variables rather than in the monomials themselves.

\subsection{'Auxiliary' scalar product for the basis states.}

The scalar product to be defined in this section is 'auxiliary'
in the sense that it does not give us the 'true' normalizations
of the basis astes of $SU(3)$.  However it is compuataionally
very convenient for us as all compuattaions with this scalar
product get reduced to simple Gaussian integrations and the
'true' normalizations themselves can then be got quite easily.  

{\bf{(i). Scalar product between generating functions of basis
states of $SU(3)$}}

We define the scalar product between any two basis states in
terms of the scalar product between the corresponding generating
functions as follows :

\begin{eqnarray}
(g', g)&=& {\int_{-\pi}^{+\pi}}{\frac{d\theta}{2\pi}} \int
\frac{d^{2}z_1}{\pi^2} \frac{d^{2}z_2}{\pi^2}
\frac{d^{2}w_1}{\pi^2} \frac{d^{2}w_2}{\pi^2}
{\exp}(-\bar{z_1}z_1 - \bar{z_2}z_2 - \bar{w_1}w_1
-\bar{w_2}w_2)\nonumber\\
&&\nonumber \\
&&\times {\exp}((r'(p'z_1+q'z_2) + s'(p'w_2-q'w_1) - \frac{-v'}{z_3}
(z_1w_1 + z_2w_2) + u'\bar{z}_3) \nonumber \\
&&\nonumber\\
&&\times {\exp}((r(pz_1 + qz_2) + s(pw_2-qw_1) - \frac{-v}{z_3}
(z_1w_1 + z_2w_2) + uz_3)\, , \nonumber \\
&&\nonumber \\
&=& (1-v'v)^{-2} \left (\sum_{n=0}^{\infty}
\frac{(u'u)^n}{(n!)^2}\right )
{\exp}\left [(1-v'v)^{-1}(p'p + q'q)(r'r + s's)\right ]\, .
\label{gg'}
\end{eqnarray}

{\bf{(ii). Choice of the variable $z_3$}}

To obtain the Eq.(\ref{gg'} we have made the choice
\begin{eqnarray}
z_3=\exp(i\theta )\, .
\label{z3}
\end{eqnarray}

The choice, Eq.(\ref{z3}), makes our basis states for $SU(3)$
depened on the variables $z_1,z_2,w_1,w_2$ and $\theta $.

{\bf{(iii). Scalar product between the gneralized generating
functions of the basis states of $SU(3)$}} 

For the generalized generating function the scalar product
becomes 

\begin{eqnarray}
{(\cal{G'}, \cal{G})} &=& (1 - v'v)^{-2} {\exp}\left [(1 - v'v)^{-1}({r_p}'r_p +
{r_q}'r_q + {s_p}'s_p + {s_q}'s_q) \right ]\nonumber \\
&&\nonumber \\
&&\times \left [{\sum_{n=0}}^{\infty} \frac{1}{(n!)^2}\left (u'- v
\frac{({r_p}'{s_q}'+ {r_q}'{s_p}')}{(1 - v'v)}\right )^n\, {\bf \cdot}\,\left (u - v'
\frac{({r_p}{s_q} + {r_q}{s_p})}{(1 - v'v)}\right )^n \right ]\,
,
\label{GIP}
\end{eqnarray}

and as in Eq.(\ref{rprqspsq})

\begin{eqnarray}
r_p=rp, \qquad r_q=rq, \qquad s_p=sp, \qquad s_q=-sq\, ,\nonumber \\
r_p'=r'p', \qquad r_q'=r'q', \qquad s_p'=s'p', \qquad
s_q'=-s'q'\, .
\label{s_q=-s_q}
\end{eqnarray}

\subsection{Normalizations}

\noindent {\bf{{(i). 'Auxiliary' normalizations of unnormalized
basis states}}} 

The scalar product between two unnormalized basis states,
computed using our 'auxiliary scalar product, is given by,

\begin{eqnarray}
M(PQRSUV)&\equiv &(PQRSUV\vert PQRSUV)\, ,\nonumber \\
&&\nonumber\\
&&=\frac{(V+P+Q+1)! }{P! Q! R! S! U! V! (P+Q+1)}\, .
\label{M}
\end{eqnarray}

\noindent {\bf{(ii). Scalar product between the unnormalized and normalized
basis states}}

The scalar product, computed using our 'auxiliary' scalar
product, between an unnormalized basis state and a normalized
one is given by the next equation where it is denoted by
$(PQRSUV\vert\vert PQRSUV>$.

\begin{equation}
(PQRSUV\vert\vert PQRSUV>=N^{-1/2}(PQRSUV)\times M(PQRSUV)
\label{(||)}
\end{equation}

\noindent {\bf{{(iii). 'True' normalizations of the basis
states}}} 

We call the ratio of the 'auxiliary' norm of the unnormalized
basis sate represented by $\vert PQRSUV)$ and the scalar product
of the unnormalized basis state with a normalized
Gelfand-Zeitlin state, represented by $\vert PQRSUV > $, as
'true' normalization.  It is given by

\begin{eqnarray}
N^{1/2}(PQRSUV)&\equiv & \frac{(PQRSUV\vert
PQRSUV)}{<PQRSUV\vert PQRSUV>}\nonumber \\ 
&&\nonumber\\
&&=\left ( \frac{(U+P+Q+1)! (V+P+Q+1)! }{P! Q! R!S! U! V!
(P+Q+1)}\right )^{1/2}\, .
\label{N}
\end{eqnarray}

\section{{{Generating function for the Wigner's $D$-matrix 
elements of $SU(3)$.}}}

\begin{equation}
g(p,q,r,s,u,v,z_1,z_2,w_1,w_2) = \sum_{P,Q,R,S,U,V}
p^Pq^Qr^Rs^Su^Uv^V\vert PQRSUV)\, ,
\end{equation}
where $\vert PQRSUV)$ is an unnormalized basis state in the IR
labeled by the two positive integers $(M=R+U, N=S+V)$.

We know from Eq.(\ref{N}),

\begin{equation}
\vert PQRSUV) = N^{(1/2)}(PQRSUV) \vert PQRSUV>\, ,
\end{equation}

where $2I=P+Q$ and $\vert PQRSUV)$ is a normalized basis state.

Therefore

\begin{equation}
g=\sum_{PQRSUV} \left (
\frac{(U+2I+1)! (V+2I+1)! } {P!Q!R!S!U!V!(2I+1)} \right )^{(1/2)} p^Pq^Qr^Rs^Su^Uv^V \vert PQRSUV>\, .
\end{equation}

Now,

\begin{eqnarray}
Ag(p,q,...)&=& \sum_{PQRSUV}\sum_{P'Q'R'S'U'V'}\left 
( \frac{(U+2I+1)!(V+2I+1)!}{P!Q!R!S!U!V!(2I+1)}\right )^{(1/2)}\nonumber \\
&& \nonumber \\
&&\times D_{PQRSUV,\,\, P'Q'R'S'U'V'}^{(M=R+U,\,\, N=S+V)}\times \,\,
p^Pq^Qr^Rs^Su^Uv^V \times \,\,\vert PQRSUV>\, .
\end{eqnarray}

To get a generating function for the matrix elements alone we
have to take the inner product of this transformed generating
function with the generating function for the basis states.
Thruoghout the following we take the variables $p, q, r, s, u,
v$ together with their primed and unprimed variants to be real
since we are interested only in the coefficients of monomials in
these different sets of varibles in different expansions and are
not interested in these variables or their functions as such. 

Thus,

\begin{eqnarray}
&&\left ( g(p", q", r", s", u", v";\,\, z_1, z_2, z_3, w_1, w_2),\,\,
A g(p, q, r, s, u, v;\,\, z_1, z_2, z_3, w_1, w_2)\right )\nonumber \\
&& \nonumber \\
&=& \sum_{PQRSUV} \sum_{P'Q'R'S'U'V'}
\sum_{P"Q"R"S"U"V"} \left
(\frac{(U+2I+1)!(V+2I+1)!}{P!Q!R!S!U!V!(2I+1)}\right
)^{(1/2)}\nonumber \\ 
&& \nonumber \\
&& \times  (P"Q"R"S"U"V"\Vert P'Q'R'S'U'V'>\times 
D_{PQRSUV,\,\, P'Q'R'S'U'V'}^{(M=R+U,\,\,N=S+V)}(A)\nonumber \\
&& \nonumber \\
&& \times p^Pq^Qr^Rs^Su^Uv^V p"^{P"}q"^{Q"}r"^{R"}s"^{S"}u"^{U"}v"^{V"}\, .
\end{eqnarray}

But we know from Eq.(\ref{(||)}),

\begin{eqnarray}
& (&P"Q"R"S"U"V" \vert\vert P'Q'R'S'U'V'>\nonumber \\
&& \nonumber \\
& = & \left (\frac{(U' + 2I' + 1)! (V' + 2I' + 1)!}{P'! Q'! R'! S'! U'! V'! (2I' + 1)}\right )^{(-1/2)}
\times \frac{(V'+P'+Q'+1)!}{P'!Q'!R'!S'!U'!V'!(P'+Q')}\nonumber \\
&& \nonumber \\
&&\times \delta_{P"P'} \delta_{Q"Q'} \delta_{R"R'} \delta_{S"S'}
\delta_{U"U'} \delta_{V"V'}\, .
\end{eqnarray}

Substituting this formula and changing the double primed
variables to single primed ones, we get

\begin{eqnarray}
&&\left ( g(p', q', r,' s', u', v';\,\, z_1,z_2,z_3,w_1,w_2){\bf ,}\quad 
 Ag(p,q,r,s,u,v;\,\, z_1,z_2,z_3,w_1,w_2) \right ) \nonumber \\
&& \nonumber \\
& = & \sum_{{\tiny {PQRSUV;\,\,P'Q'R'S'U'V'}}} \left
(\frac{(U+2I+1)!(V+2I+1)!} {P! Q! R! S! U! V! (2I +1)})\times
 (\frac{P'! Q'! R'! S'! U'! V'! (2I' + 1)}{(U'+2I'+1)!(V'+2I'+1)!}\right )^{(1/2)}
\nonumber \\
&&\nonumber \\
&&\times (\frac{(V'+P'+Q'+1)!}{P'!Q'!R'!S'!U'!V'!(P'+Q'+1)})
\times D_{PQRSUV,\,\,P'Q'R'S'U'V'}^{(M=R+U,\,\,N=S+V)}(A) \nonumber \\
&&\nonumber \\
&& \times p^P q^Q r^R s^S u^U v^V{p'}^{P'}{q'}^{Q'}{r'}^{R'}{s'}^{S'}{u'}^{U'}{v'}^{V'}\, .
\end{eqnarray}

We therefore conclude that the Wigner's $D$-matrix element,

\begin{eqnarray*}
D_{PQRSUV,\,\, P'Q'R'S'U'V'}^{(M = R + U, \,\,N = S + V)}\, 
\end{eqnarray*}

for $SU(3)$ can be obtained as the coefficient of the monomial,

\begin{eqnarray*}
p^P q^Q r^R s^S u^U v^V \times {p'}^{P'} {q'}^{Q'} {r'}^{R'}
{s'}^{S'} {u'}^{U'} {v'}^{V'}\, ,
\end{eqnarray*}

multiplied by,

\begin{eqnarray}
\left ( \frac{P! Q! R! S! U! V! (2I + 1)}{(U + 2I + 1)! (V +  2I + 1)! }
\times \frac{(U' + 2I' + 1)! (V' + 2I' + 1)!}{P'! Q'! R'! S'! U'! V'! (2I' + 1)} 
 \right )^{(1/2)}\nonumber \\
\nonumber \\
\times (\frac{P'!Q'!R'!S'!U'!V'!(P'+Q'+1)}{(V'+P'+Q'+1)!})\, ,
\label{MFDM}
\end{eqnarray}
in the inner product $(g',\,\, Ag)$ between the untransformed and
transformed generating functions for the basis states.

Next we calculate this inner product using the explicit
realization for the generating function.  For this purpose it is
advantageous, as will be seen in a minute, to use the
generalized generating function for the basis states

\begin{eqnarray}
{\cal G} &=& {\exp}(r_pz_1+r_qz_2+s_pw_2+s_qw_1+uz_3+vw_3)\nonumber\\ 
&& \nonumber \\
& =& {\exp}\left ( {\pmatrix{r_p & r_q & u}}{\pmatrix{z_1\cr z_2 \cr z_3}} +
{\pmatrix{w_1 & w_2 & w_3}}{\pmatrix{s_q \cr s_p \cr v}}\right
)\, .
\end{eqnarray}

When any element $A \in SU(3)$ acts on this generating function it undergoes
the following transformation

\begin{eqnarray}
A{\cal G} = {\exp}\left ( {\pmatrix{r_p & r_q & u}} A{\pmatrix{z_1\cr z_2 \cr z_3}} +
{\pmatrix{w_1 & w_2 & w_3}}A^{\dagger} {\pmatrix{s_q \cr s_p \cr
v}}\right )\, .
\end{eqnarray}

As is clear from the above equation we can let the triplets
$r_p, r_q, u$ and $s_q,s_p,v$ undergo the transformation instead
of the triplets $z_1,z_2,z_3$ and $w_1,w_2,w_3$.  Therfore we
can write the transformed generating function as

\begin{eqnarray}
A{\cal G} = {\cal G}(r_p",r_q",u"; s_q",s_p",v")\, ,
\end{eqnarray}

where

\begin{eqnarray}
r_p"&=& a_{11}r_p+a_{21}r_q+a_{31}u\nonumber \\
r_q"&=& a_{12}r_p+a_{22}r_q+a_{32}u\nonumber \\
u"&=& a_{13}r_p+a_{23}r_q+a_{33}u\, ,\nonumber \\
&&  \nonumber \\
s_q"&=& a^*_{11}s_q+a^*_{21}s_p+a^*_{31}v\nonumber \\
s_p"&=& a^*_{12}s_q+a^*_{22}s_p+a^*_{32}v\nonumber \\
v"&=& a^*_{13}s_q+a^*_{23}s_p+a^*_{33}v\, .
\label{A*GGF}
\end{eqnarray}



To continue with our computation we have to take the inner
product of this transformed generating function with the
(untransformed) generating function of the basis states.

This is known to us from Eq.(\ref{GIP}) as

\begin{eqnarray}
{(\cal{G'}, \cal{G"})} &=& (1 - v'v")^{-2} {\exp} \left [ (1 - v'v")^{-1}({r_p}'r_p" +
{r_q}'r_q" + {s_p}'s_p" + {s_q}'s_q") \right ] \nonumber \\
&& \nonumber \\
&& \times \left [ {\sum^{\infty}_{n=0}}\frac{1}{(n!)^2}\left (u'- v"
\frac{({r_p}'{s_q}'+ {r_q}'{s_p}')}{(1 - v'v")}\right )^n \left (u" - v'
\frac{({r_p}"{s_q}" + {r_q}"{s_p}")}{(1 - v'v")}\right )^n \right ]\, .
\end{eqnarray}

This expression gets further simplified if we substitute from
Eq.(\ref{rprqspsq})

\begin{eqnarray*}
r_p'=r'p', \quad r_q'=r'q', \quad s_q'=-s'q', \quad s_p'=s'p\, .
\end{eqnarray*}

We, therefore, get

\begin{eqnarray}
{(\cal{G'}, \cal{G"})} &=& (1 - v'v")^{-2} {\exp} \left [ (1 - v'v")^{-1}({r_p}'r_p" +
{r_q}'r_q" + {s_p}'s_p" + {s_q}'s_q") \right ] \nonumber \\
&& \nonumber \\
&&\times \left [ {\sum^{\infty}_{n=0}} \frac{1}{(n!)^2}(u')^n (u" - v'
\frac{({r_p}"{s_q}" + {r_q}"{s_p}")}{(1 - v'v")})^n \right ]\, .
\label{GG'}
\end{eqnarray}

One last simplification can be brought about in the above expression
when we recognize that

\begin{eqnarray}
{r_p}"{s_q}" + {r_q}"{s_p}"+u"v" = & {r_p}{s_q} +
{r_q}{s_p}+vu\, ,\nonumber\\
 = & vu\, .
\end{eqnarray}

This tells us that

\begin{eqnarray}
 {r_p}"{s_q}" + {r_q}"{s_p}"= uv - u"v"\, .
\end{eqnarray}

Substituting this in our expression Eq.(\ref{GG'}) for the inner
product we get,

\begin{eqnarray}
{(\cal{G'}, \cal{G"})} &=& (1 - v'v")^{-2} {\exp} \left [ (1 -
v'v")^{-1}({r_p}'r_p" + {r_q}'r_q" + {s_p}'s_p" + {s_q}'s_q")
\right ] \nonumber \\
&& \nonumber \\
&& \times \left [ {\sum^{\infty}_{n=0}} \frac{1}{(n!)^2}(u')^n
(u" - v'
\frac{({u}{v} - {u}"{v}")}{(1 - v'v")})^n \right ]\nonumber \\
&& \nonumber \\
& =& (1 - v'v")^{-2} {\exp} \left [ (1 - v'v")^{-1}({r_p}'r_p" +
{r_q}'r_q" + {s_p}'s_p" + {s_q}'s_q") \right ] \nonumber \\ 
&& \nonumber \\
&&\times \left [ {\sum^{\infty}_{n=0}} \frac{1}{(n!)^2}\left ( {u'} 
\frac{({u}" - {u}{v}v')}{(1 - v'v")} \right )^n\right ]\, .
\label{GFWD1}
\end{eqnarray}

On the other hand if we use our present slightly modified scalar
product then,

\begin{eqnarray}
{(\cal{G'}, \cal{G"})} 
&=& (1 - v'v")^{-2} {\exp} 
\left [ 
\frac{({r_p}'r_p" +{r_q}'r_q" + {s_p}'s_p" + {s_q}'s_q")}{(1 -
v'v")} \right. \nonumber \\ 
&&\nonumber \\ 
&&+ \left.  \left ( u'- v"\frac{({r_p}'{s_q}' + {r_q}'{s_p}')}{(1 - v'v")}\right ) 
            \left (u" - v'\frac{({r_p}"{s_q}" + {r_q}"{s_p}")}{(1 - v'v")}\right ) 
\right ]\, ,\nonumber \\
&&\nonumber \\
&=& (1 - v'v")^{-2} {\exp} \left [ 
\frac{({r_p}'r_p" +{r_q}'r_q" + {s_p}'s_p" + {s_q}'s_q")
+u'(u" - uvv')}{(1 - v'v")}\right ]
\label{GFWD2}
\end{eqnarray}

The expression on the right hand side of Eq.(\ref{GFWD1}) or of
Eq.(\ref{GFWD2}) is our generating function for the Wigner's
$D$-matrix elements of $SU(3)$.

\section{{{Wigner's $D$-matrix elements of $SU(3)$ in
any irreducible representation.}}}

In this section our task is to extract the coefficient of the mononial

\begin{eqnarray*}
p^P q^Q r^R s^S u^U v^V 
\times {p'}^{P'} {q'}^{Q'} {r'}^{R'} {s'}^{S'} {u'}^{U'} {v'}^{V'}\, .
\end{eqnarray*}
in the expansion of the generating function that we have
obtained above, Eq.(\ref{GFWD1}), for the Wigner's $D$-matrix
elements of $SU(3)$.  For this purpose we expand the right hand
side of the above generating function and obtain

\begin{eqnarray}
&\sum ^\infty_{m=0}& {\frac{({r_p}'r_p" + {r_q}'r_q" + {s_p}'s_p"
+ {s_q}'s_q")}{m!(1 - v'v")^{m}}} \times  {\sum^{\infty}_{n=0}}
\frac{1}{(n!)^2}\left ( {u'}
\frac{({u}" - {u}{v}v')}{(1 - v'v")^{(1+2/n)}} \right )^{n}\, ,\nonumber \\
&& \nonumber \\
&=&\!\sum^\infty_{m=n=s=0}\,\,\sum^{n,m,m-m_1,m-m_1-m_2}_{t,m_1,m_2,m_3=0}
{\frac{(\!s+m+n+1\!)!}{m! n! (\!n-t\!)! t!  (\!m+n+1\!)!
s!(m-m_1-m_2-m_3)! m_3!}}\nonumber \\
&& \nonumber \\
&&\times {(r_p")^{m_1}(r_q")^{m_2}(s_p")^{m_3} (s_q")^{m-\sum m_i}}
(p')^{m_1+m_3} (q')^{m-m_1-m_3} (r')^{m_1+m_2} (s')^{m-m_1-m_2}\nonumber \\ 
&& \nonumber \\
&&\times (u')^n (v')^{n-t+s} u'^nv'^{n-t+s}u"^tv"^s (-uv)^{n-t}\, . \nonumber \\
\end{eqnarray}

Now let,

\begin{eqnarray}
m_1+m_2    &=&P'\, ,\nonumber \\
m  -m_1-m_3&=&Q'\, ,\nonumber \\
m_1+m_2    &=&R'\, ,\nonumber \\
m  -m_1-m_2&=&S'\, ,\nonumber \\
n          &=&U'\, ,\nonumber \\
n  -t  +s  &=&V' \, .
\end{eqnarray}

The above assignments imply,

\begin{eqnarray}
m      &=&P'+Q'\, \nonumber \\
m_2-m_3&=&R'-P'\, \nonumber \\
m_2    &=&R'-P'+m_3\, ,\nonumber \\
{\mbox {and,~~~~~~~~~~~~~~~~ }} s&=&t+V'-U'\, .
\end{eqnarray}

This gives us

\begin{eqnarray}
&& D_{PQRSUV,\,\, P'Q'R'S'U'V'}^{(M = R + U,\,\, N = S + V)}(A)\nonumber \\
&& \nonumber \\
&=&\sum^\infty_{P'+Q'=0} \sum^\infty_{U'=0} \sum^\infty_{V'=U'}
\sum^{P'+Q}_{m_1=0} \sum^{U'}_{t=0} \sum^{S'}_{m_3=0}
\sum^{P'+Q'-m_1}_{m_2=0}\nonumber \\
&& \nonumber \\
&&\times {\frac{(t+V'-U')! (-uv)^{U'-t}}{(P'+Q')! U'! (U'-t)! t!
(P'+Q'+U'+1)!  (t+V'-U')! (R'-P'+m_3)! (S'-m_3)! m_3!}}\nonumber\\
&& \nonumber \\
&&\times {(r_p")^{m_1}(r_q")^{m_2}(s_p")^{m_3}
(s_q")^{S'-m_3}}u"^tv"^{t+V'-U'}
\times (p')^{P'}(q')^{Q'} (r')^{R'} (s')^{S'} (u')^{U'}
(v')^{V'}\, .\nonumber\\
\end{eqnarray}

In the above we substitute for the following from Eq.(\ref{A*GGF})

\begin{eqnarray*}
r_p", \quad r_q", \quad u", \quad s_q", \quad s_p", \quad v" \, ,
\end{eqnarray*}

and get, 

\begin{eqnarray}
&& D_{PQRSUV,\,\, P'Q'R'S'U'V'}^{(M = R + U,\,\, N = S + V)}(A)\nonumber \\
&& \nonumber \\
&=&\sum^\infty_{P'+Q'=0} \sum^\infty_{U'=0} \sum^\infty_{V'=U'}
\sum^{P'+Q}_{m_1=0} \quad \sum^{U'}_{t=0} \quad \sum^{S'}_{m_3=0}
\sum^{P'+Q'-m_1}_{m_2=0}
\sum_{m_{11}+m_{12}+m_{13}=m_1} \quad \sum_{m_{21}+m_{22}+m_{23}=m_2}\nonumber \\
&& \nonumber \\
&& \sum_{m_{31}+m_{32}+m_{33}=m_3}
\sum_{m_{41}+m_{42}+m_{43}=S'-m_3}\quad \sum_{t_{11}+m_{12}+m_{13}=t}
\quad \sum_{t_{21}+t_{22}+t_{23}=t+V'-U'}\nonumber \\
&& \nonumber \\
&&{\frac{(-1)^{U'-t}(t+V'-U')! (-uv)^{U'-t}}{(P'+Q')! U'! (U'-t)! t!
(P'+Q'+U'+1)!  (t+V'-U')! (R'-P'+m_3)! (S'-m_3)! m_3!}}\nonumber\\ 
&& \nonumber \\
&&\times {\frac{m_1!m_2!m_3!(S'-m_3)! t!
(t+V'-U')!}{m_{11}!m_{12}!m_{13}!  m_{21}!m_{22}!m_{23}!
m_{31}!m_{32}!m_{33}!  m_{41}!m_{42}!m_{43}!
t_{11}!t_{12}!t_{13}! t_{21}!t_{22}!t_{23}!}}\nonumber \\
&& \nonumber \\
&&\times (a_{11})^{m_{11}} (a_{11}^*)^{m_{11}} (a_{21})^{m_{12}}
(a_{21}^*)^{m_{42}} (a_{31})^{m_{13}} (a_{31}^*)^{m_{43}}
(a_{12})^{m_{21}} (a_{12}^*)^{m_{31}}
(a_{22})^{m_{22}} (a_{22}^*)^{m_{32}} (a_{32})^{m_{23}}\nonumber \\
&& \nonumber \\
&&\times (a_{32}^*)^{m_{33}} (a_{13})^{t_{11}} (a_{13}^*)^{t_{21}}
(a_{23})^{t_{12}} (a_{23}^*)^{t_{22}}
(a_{33})^{t_{13}} (a_{33}^*)^{t_{23}}\nonumber \\ 
&& \nonumber \\
&&\times (p')^{P'}(q')^{Q'} (r')^{R'} (s')^{S'} (u')^{U'} (v')^{V'} 
\times (p)^{P} (q)^{Q} (r)^{R} (s)^{S} (u)^{U} (v)^{V}\, .
\end{eqnarray}
where we have made the identifications,

\begin{eqnarray}
m_{11}+m_{21}+t_{11}+m_{32}+m_{42}+t_{22}&=&P \, ,\nonumber \\
m_{12}+m_{22}+t_{12}+m_{41}+m_{32}+t_{22}&=&Q \, ,\nonumber \\
m_{11}+m_{21}+t_{11}+m_{12}+m_{22}+t_{22}&=&R \, ,\nonumber \\
m_{41}+m_{31}+t_{21}+m_{42}+m_{32}+t_{22}&=&S \, ,\nonumber \\
m_{13}+m_{23}+t_{13}+U'-t&=&U \nonumber \, ,\\
m_{43}+m_{33}+t_{23}+U'-t&=&V\, .\nonumber\\
\end{eqnarray}

Finally, we get the desired object i.e., the Wigner's $D$-matrix
or the finite transformation matrix of the group $SU(3)$ in any
irreducible representation by multiplying the above matrix
element by the factor in Eq.(\ref{MFDM}).  

So finally,

\begin{eqnarray}
&&D_{PQRSUV,\,\, P'Q'R'S'U'V'}^{(M = R + U,\,\, N = S + V)}(A)\nonumber \\
&& \nonumber \\
&=&\left ( \frac{P! Q! R! S! U! V! (2I + 1)}{(U + 2I + 1)! (V +  2I + 1)! }
\times \frac{(U' + 2I' + 1)! (V' + 2I' + 1)!}{P'! Q'! R'! S'! U'! V'! (2I' + 1)} 
 \right )^{(1/2)}\nonumber \\
&&\nonumber \\
&&\times (\frac{P'!Q'!R'!S'!U'!V'!(P'+Q'+1)}{(V'+P'+Q'+1)!}) \nonumber \\
&&\nonumber \\
&&\times \sum^\infty_{P'+Q'=0} \sum^\infty_{U'=0} \sum^\infty_{V'=U'}
\sum^{P'+Q}_{m_1=0} \sum^{U'}_{t=0} \sum^{S'}_{m_3=0}
\sum^{P'+Q'-m_1}_{m_2=0}
\sum_{m_{11}+m_{12}+m_{13}=m_1} \quad \sum_{m_{21}+m_{22}+m_{23}=m_2}\nonumber \\
&& \nonumber \\
&& \sum_{m_{31}+m_{32}+m_{33}=m_3}
\quad \sum_{m_{41}+m_{42}+m_{43}=S'-m_3} \quad \sum_{t_{11}+m_{12}+m_{13}=t}
\quad \sum_{t_{21}+t_{22}+t_{23}=t+V'-U'}\nonumber \\
&& \nonumber \\
&&\times {\frac{(-1)^{U'-t}(t+V'-U')! (-uv)^{U'-t}}{(P'+Q')! U'! (U'-t)! t!
(P'+Q'+U'+1)!  (t+V'-U')! (R'-P'+m_3)! (S'-m_3)! m_3!}}\nonumber\\ 
&& \nonumber \\
&&\times {\frac{m_1!m_2!m_3!(S'-m_3)! t!
(t+V'-U')!}{m_{11}!m_{12}!m_{13}!  m_{21}!m_{22}!m_{23}!
m_{31}!m_{32}!m_{33}!  m_{41}!m_{42}!m_{43}!
t_{11}!t_{12}!t_{13}! t_{21}!t_{22}!t_{23}!}}\nonumber \\
&& \nonumber \\
&&\times (a_{11})^{m_{11}} (a_{11}^*)^{m_{11}} (a_{21})^{m_{12}}
(a_{21}^*)^{m_{42}} (a_{31})^{m_{13}} (a_{31}^*)^{m_{43}}
(a_{12})^{m_{21}} (a_{12}^*)^{m_{31}}
(a_{22})^{m_{22}} (a_{22}^*)^{m_{32}} (a_{32})^{m_{23}}\nonumber \\
&& \nonumber \\
&&\times (a_{32}^*)^{m_{33}} (a_{13})^{t_{11}} (a_{13}^*)^{t_{21}}
(a_{23})^{t_{12}} (a_{23}^*)^{t_{22}}
\left. (a_{33})^{t_{13}} (a_{33}^*)^{t_{23}} \right )\, .
\label{WDME}
\end{eqnarray}

The above equation, Eq.(\ref{WDME}) for the Wigner's $D$-matrix
element for $SU(3)$ is the analogue of Wigner's $D$-matrix
element for $SU(2)$(see for example {\cite{tjd,sj}}).

\section{Discussion.}

In this paper, making use of the tools of a calculus that we had
set up previously to do computations on $SU(3)$, we have
obtained (i) a generating function(Eq.(\ref{GFWD1}),(\ref{GFWD2})) for the
Wigner's $D$-matrix elements of $SU(3)$ and (ii) a closed form
algebraic expression(Eq.(\ref{WDME})) for the individual
Wigner's $D$-matrix elements of $SU(3)$ in any irreducible
representation.  To our knowledge this is the first time that a
such generating function has been written for $SU(3)$. But this
generating function gives us unitary matrix elements of $SU(3)$
only up to a multiplicative factor.  The reason for this is that
our auxiliary measure for the basis states is not a group
invariant measure.  This is clearly a drawback.  However for
computing objects such as the group characters this is no hurdle
since the characters are invariant under basis transformtions.

We also note that our generating function is in fact a product
of two factors one of which is an exponential function and the
second is some power series.  This seems to be a consequence of
the particular choice of variables occuring in the construction
of our basis functions which makes it possible for the $\theta$
variable part of any object of interest, such as for example the
Clebsch-Gordan coefficients etc, to decouple from the part that
dependeds on other variables.  Next, the expression for the
individual $D$-matrix elements for $SU(3)$ has been obtained by
many people previously also \cite{cemm,ntj,hdf}.  But one
desirable feature about our expression is that it is quite
compact and is independent of any particular parametrization
used for describing the defining representation of $SU(3)$.  Now
coming back to the generating function for the matrix elements
we recall, from our experience in computing the Clebsch-Gordan
coefficients of $SU(3)$ previously and now the present
computation of $D$-matrix elements that problems which are
intractable by other methods may be, some times, easier to deal
with using the generating function method.  Therefore it is
hoped that, now that a calculus and a generating function for
Wigner's $D$-matrix elements are available, one might be able to
employ this technique profitably to problems of interest in some
areas of physics.

\renewcommand{\theequation}{A.\arabic{equation}}
\setcounter{equation}{0}

\newpage

{\large\bf Appendix : Examples}
To compute the matrix elements of $SU(3)$, for lower dimensions,
it is easier to work with the generating function for the matrix
elements Eq.(\ref{GFWD1}), (\ref{GFWD2})).  

For the irreducible representation $\underline 3$ the only terms
of the generating function which are relavent are the ones
linear in the primed and and doubly primed composite variables
$r_p', r_p^{''} \cdots $.  This gives us the following expansion
for the generating function

\begin{eqnarray}
r_p'r_p^{''} + r_q'r_q^{''}+ s_p's_p^{''}+
s_q's_q^{''}+u'u^{''}+v'v^{''}\, .
\end{eqnarray}

We now substitute for the doubly primed variables, in the above
expression, from the Eqs.((\ref{A*GGF}), (\ref{s_q=-s_q})), and extract the
coefficients of the various monomials $p^Pq^Qr^Rs^Su^Uv^V$ for
the values of the quantum numbers $P,Q,R,S,U,V$ given in the
table below for the IR $\underline 3$.  This gives us the $SU(3)$
representative matrix Eq(\ref{A}).

\vspace{1cm}

\begin{center}
$\underline{3}$($M$=1, $N$=0)\\
\vspace{0.5cm}
\begin{tabular}{|c|c|c|c|c|c|c|c|c|c|c|c|}\hline
. & $P$ & $Q$ & $R$ & $S$ & $U$ & $V$ & $I$   & $I_3$  & $Y$ & $\vert PQRSTUV)$ & $N^{1/2}$ \\ \hline 
$u$ & 1 & 0 & 1 & 0 & 0 & 0 & 1/2 & $1/2$  & 1/3   & $z_1$            & $\sqrt{2}$ \\ \hline 
$d$ & 0 & 1 & 1 & 0 & 0 & 0 & 1/2 & $-1/2$ & 1/3   & $z_2$            & $\sqrt{2}$ \\ \hline 
$s$ & 0 & 0 & 0 & 0 & 1 & 0 & 0   &  0   & -2/3  & $z_3$            & $\sqrt{2}$ \\ \hline 
\end{tabular}
\end{center}

\vspace{1cm}

\begin{center}
$\underline{3}$($M$=1, $N$=0)\\
\vspace{0.5cm}
\begin{tabular}{|c|c|c|c|}\hline
.    & $u$             & $d$             & $s$      \\ \hline 
$u$  & $a_{11}$        & $a_{12}$        & $a_{13}$ \\ \hline 
$d$  & $a_{21}$        & $a_{22}$        & $a_{23}$ \\ \hline 
$s$  & $a_{31}$        & $a_{32}$        & $a_{33}$ \\ \hline 
\end{tabular}
\end{center}

A similar treatment for the IR $\underline 3^*$, using the
corresponding table, given below, gives us the $SU(3)$ matrix
$A^\dagger$.

\vspace{1cm}

\begin{center}
$\underline{3}^*$($M$=0, $N$=1)\\
\vspace{0.5cm}
\begin{tabular}{|c|c|c|c|c|c|c|c|c|c|c|c|}\hline
.         & $P$ & $Q$ & $R$  & $S$  & $U$ & $V$ & $I$   & $I_3$  & $Y$    & $\vert PQRSTUV)$ & $N^{1/2}$ \\ \hline 
$\bar{d}$ & 1   & 0   &  0   &  1   &  0 & 0 &   1/2    & $1/2$  & -1/3   & $w_2$            & $\sqrt{2}$ \\ \hline 
$\bar{u}$ & 0   & 1   &  0   &  0   &  0 & 0 &   1/2    & $-1/2$ & -1/3    & -$w_1$            & $\sqrt{2}$ \\ \hline 
$\bar{s}$ & 0   & 0   &  0   &  0   &  0 & 1 &   0      &  0     & 2/3      & $w_3$            & $\sqrt{2}$ \\ \hline 
\end{tabular}
\end{center}

\vspace{1cm}

\begin{center}
$\underline{3}^*$($M$=0, $N$=1)\\
\vspace{0.5cm}
\begin{tabular}{|c|c|c|c|}\hline
.                & $\bar{d}$         & $\bar{u}$         & $\bar{s}$ \\ \hline 
$\bar{d}$        & $a_{11}^*$        & $a_{21}^*$        & $a_{31}^*$        \\ \hline 
$\bar{u}$        & $a_{12}^*$        & $a_{22}^*$        & $a_{32}^*$        \\ \hline 
$\bar{s}$        & $a_{13}^*$        & $a_{23}^*$        & $a_{33}^*$        \\ \hline 
\end{tabular}
\end{center}

We now treat the case of the IR $\underline 8$.  The terms
relevant for this IR are quadratic in the primed and doubly
primed composite variables.  For example the first term in the
expansion of the generating function Eq.(\ref{GFWD2}) is

\begin{eqnarray}
r_p'r_p^{''}s_p's_p^{''}&=& p'^2r's'\left ( -A_{11}A^*_{11}pqrs
+ A_{11}A^*_{12}A^*_{12}p^2rs + A_{11}A^*_{13}prv - A_{21}A^*_{12}
q^2rs\right. \nonumber \\ 
\nonumber \\
&&\left.  + A_{21}A^*_{12}pqrs + A_{21}
A^*_{13} qrv + A_{31}A^*_{11} qsu + A_{31}A^*_{12} psu +
A_{31}A^*_{13}uv\right )\, .\nonumber\\
\label{8*8row1}
\end{eqnarray}

In Eq.(\ref{8*8row1}) the various monomials $p^Pq^Qr^Rs^Su^Uv^V$
correspond to the quantum numbers $P,Q,R,S,U,V$ in the first row
of the table corresponding to the IR $\underline 8$ as indicated
in the table given below.  Therefore their coefficients give us
the first row of the $SU(3)$ Wigner's $D$-matrix for the IR
$\underline 8$.  One can build the remaining rows in a similar
manner.  The result is given in the form of a table below.

\vspace{1cm}

\begin{center}
$\underline{8}$($M$=1, $N$=1)\\
\vspace{0.3cm}
\begin{tabular}{|c|c|c|c|c|c|c|c|c|c|c|c|}\hline
.            & $P$ & $Q$ & $R$ & $S$ & $U$ & $V$ & $I$  & $I_3$  & $Y$ & $\vert PQRSTUV)$ & $N^{1/2}$   \\ \hline 
$\pi^+$      & 2   & 0   &  1  &  1  &  0  & 0   &  1   &  1     & 0   & $z_1w_2$         & $\sqrt{6}$  \\ \hline 
$\pi^0$      & 1   & 1   &  1  &  1  &  0  & 0   &  1   &  0     & 0   & $-z_1w_1+z_2w_2$ & $\sqrt{12}$ \\ \hline 
$\pi^-$      & 0   & 2   &  1  &  1  &  0  & 0   &  1   & -1     & 0   & $-z_2w_1$        & $\sqrt{6}$  \\ \hline 
$K^+$        & 1   & 0   &  1  &  0  &  0  & 1   &  1/2 &  1/2   & 1   & $z_1w_3$         & $\sqrt{6}$  \\ \hline
$K^0$        & 0   & 1   &  1  &  0  &  0  & 1   &  1/2 &  -1/2  & 1   & $z_2w_3$         & $\sqrt{6}$  \\ \hline
$\bar{K}^0 $ & 1   & 0   &  0  &  1  &  1  & 0   &  1/2 &  1/2   & -1  & $w_2z_3$	  & $\sqrt{6}$  \\ \hline
$K^-$        & 0   & 1   &  0  &  1  &  1  & 0   &  1/2 &  -1/2  & -1  & $-w_1z_3$ 	  & $\sqrt{6}$  \\ \hline
$\eta $      & 0   & 0   &  0  &  0  &  1  & 1   &  0   &  0     & 0   & $(z_3w_3=-z_1w_1-z_2w_2)$ & 2   \\ \hline
\end{tabular}
\end{center}

\vspace{1cm}

\begin{center}
$\underline{8}^*$($M$=1, $N$=1)\\
\vspace{0.3cm}
\begin{tabular}{|c|c|c|c|c|c|c|c|c|c|}\hline
.           & $\pi^+$  & $\pi^0$  & $\pi^-$ & $K^+$ & $K^0$ & $\bar{K}^0$ & $K^-$ & $\eta$  \\ \hline 
$\pi^+$     & $(a_{21}a^*_{12}-a_{11}a^*_{11})$       & $\frac{a_{11}a^*_{12}}{\sqrt{2}}$          & $a_{11}a^*_{13}$      & $\frac{-\sqrt{2}a_{21}a^*_{12}}{3}$    & $\frac{\sqrt{2}a_{21}a^*_{13}}{3}$    & $-\sqrt{2}a_{31}a^*_{11}$          & $\sqrt{2}a_{31}a^*_{12}$    & $\frac{a_{31}a^*_{13}}{\sqrt{3}}$       \\ \hline
$\pi^0$     & $\frac{(a_{21}a^*_{21}-a_{11}a^*_{11})}{\sqrt{2}}$       & $\frac{a_{11}a^*_{21}}{2}$          & $\frac{a_{11}a^*_{31}}{\sqrt{2}}$      & $-\frac{a_{21}a^*_{11}}{6\sqrt{2}}$    & $\frac{a_{21}a^*_{31}}{6\sqrt{2}}$   & $-\frac{a_{31}a^*_{11}}{\sqrt{2}}$          & $\frac{a_{31}a^*_{21}}{\sqrt{2}}$    & $\frac{a_{31}a^*_{31}}{2\sqrt{3}}$     \\ \hline     
$\pi^-$     & $(a_{22}a^*_{21}-a_{12}a^*_{11})$       & $\frac{a_{12}a^*_{21}}{\sqrt{2}}$          & $a_{12}a^*_{31}$      & $-\frac{\sqrt{2}a_{22}a^*_{11}}{3}$    & $\frac{\sqrt{2}a_{22}a^*_{31}}{3}$    & $-\sqrt{2}a_{32}a^*_{11}$          & $\sqrt{2}a_{32}a^*_{21}$    & $\frac{a_{32}a^*_{31}}{\sqrt{3}}$  \\ \hline 
$K^+$       & $(a_{21}a^*_{23}-a_{11}a^*_{13})$       & $\frac{a_{11}a^*_{23}}{\sqrt{2}}$          & $\frac{a_{11}a^*_{33}}{\sqrt{2}}$      & $-\frac{a_{21}a^*_{13}}{3}$    & $\frac{a_{21}a^*_{33}}{3}$   & $-a_{31}a^*_{13}$          & $a_{31}a^*_{23}$    & $\frac{a_{31}a^*_{33}}{\sqrt{6}}$ \\ \hline
$K^0$       & $(a_{21}a^*_{23}-a_{12}a^*_{13})$       & $\frac{a_{12}a^*_{23}}{\sqrt{2}}$          & $a_{12}a^*_{33}$      & $-\frac{a_{21}a^*_{13}}{3}$    & $\frac{a_{21}a^*_{33}}{3}$   & $-a_{31}a^*_{13}$          & $a_{31}a^*_{23}$    & $\frac{a_{31}a^*_{33}}{\sqrt{6}}$ \\ \hline
$\bar{K}^0$ & $(a_{23}a^*_{22}-a_{13}a^*_{12})$       & $\frac{a_{13}a^*_{22}}{\sqrt{2}}$          & $a_{13}a^*_{32}$      & $-\frac{a_{23}a^*_{12}}{3}$    & $\frac{a_{23}a^*_{32}}{3}$   & $-a_{33}a^*_{12}$          & $a_{33}a^*_{22}$    & $\frac{a_{33}a^*_{32}}{\sqrt{6}}$ \\ \hline
$K^-$       & $(a_{23}a^*_{21}-a_{13}a^*_{11})$       & $\frac{a_{13}a^*_{21}}{\sqrt{2}}$          & $a_{13}a^*_{31}$      & $-\frac{a_{23}a^*_{11}}{3}$    & $\frac{a_{23}a^*_{31}}{3}$   & $-a_{33}a^*_{11}$          & $a_{33}a^*_{21}$    & $\frac{a_{33}a^*_{31}}{\sqrt{6}}$\\ \hline
$\eta$      & $\frac{\sqrt{3}(a_{23}a^*_{23}-a_{13}a^*_{13})}{\sqrt{2}}$       & $\frac{\sqrt{3}a_{13}a^*_{23}}{2}$          & $\frac{\sqrt{3}a_{13}a^*_{33}}{\sqrt{2}}$      & $-\frac{a_{23}a^*_{13}}{\sqrt{6}}$    & $\frac{a_{23}a^*_{33}}{\sqrt{6}}$   & $-\frac{\sqrt{3}a_{33}a^*_{13}}{\sqrt{2}}$          & $\frac{\sqrt{3}a_{33}a^*_{23}}{\sqrt{2}}$    & $\frac{a_{33}a^*_{33}}{2}$ \\ \hline
\end{tabular}
\end{center}

In all the above computations a normalization factor for each
$D$-matrix element is computed with the help of the Eq.(\ref{MFDM}).

\end{document}